\documentstyle[twocolumn,prl,aps]{revtex}
\begin{document}
\draft
\preprint{  }
\widetext
\title{Bloch oscillations, Zener tunneling and Wannier-Stark
ladders in the time-domain}
\author{Jon Rotvig$ ^a$, Antti-Pekka Jauho$ ^b$, and Henrik Smith$ ^a$}
\address{
$ ^a${\O}rsted Laboratory, H.C.{\O}rsted Institute,
Universitetsparken 5, University
of Copenhagen, DK-2100 Copenhagen {\O}, Denmark}
\address{
$ ^b$MIC, Technical University of Denmark, DK-2800 Lyngby, Denmark}
\date{\today}
\maketitle

\begin{abstract}
We present a time-domain analysis of carrier dynamics in a
semiconductor superlattice with two minibands.
Integration of the density-matrix equations of motion reveals
a number of new features: (i) for certain values of the applied
static electric field strong interband transitions occur;
(ii) in static fields the complex time-dependence
of the density-matrix displays
a sequence of stable plateaus in the low field regime,
and (iii) for applied fields with a periodic time-dependence the
dynamic response can be understood in terms of the quasienergy spectra.
\end{abstract}

\narrowtext
Bloch oscillations (BO) are one of the most striking predictions of the
semiclassical theory of electronic transport: in any system of
independent electrons in a periodic potential the electron
velocity becomes a periodic function of time with characteristic
frequency $\omega_B=eEd/\hbar$, where $d$ is the lattice period
and $E$ is the applied field \cite{Bloch,Zener}.
In ordinary bulk materials these oscillations cannot be seen, because
collisions dephase the coherent motion of electrons on a time-scale
which is much shorter than $T_B=2\pi/\omega_B$.  However, as pointed out
by Esaki and Tsu \cite{EsakiTsu}, the conditions for observing
BO's are much less stringent for high-quality semiconductor superlattices.
Recent years have witnessed an intense experimental activity in this
area, culminating in the observation of terahertz radiation from
coherently oscillating electrons \cite{Waschke}.

There has been equal activity on the theoretical side.
Holthaus\cite{Holthaus} analyzed the semiclassical
motion of electrons in a single mini-band subjected
to a strong {\it alternating} electric field.
Studies of this kind have gained importance due
to the emerging
free-electron lasers, which open the possibility of
experimental probing of the theoretical
predictions.  For certain values of the system
parameters a dynamical localization takes place
\cite{Dunlap}: the average velocity vanishes.  This
phenomenon can alternatively be called
band collapse \cite{Holthaus}.
Ignatov {\it et al.} \cite{Ignatov} pointed out an interesting
analog between Bloch electrons and the Josephson effect.
Very recently, Meier {\it et al.} \cite{Meier} considered coherent
motion of photoexcited carriers in the presence of
Coulomb interaction, and found out that BO's should persist
even in the limit where the exciton binding energy is
comparable to the miniband width.

The papers quoted above have mainly concentrated on
studying systems with one miniband \cite{Holt2B}; the central theme
in the present work is to
study the dynamics of electrons
in a {\it two-band} superlattice. The second miniband
adds an essential feature to the model: it is possible
to study how Zener tunneling affects the
dynamics of the carriers. Our method
consists of setting up, and solving, the density-matrix
equations of motion for the two-band system.  In this paper we focus
on the coherent part of the motion.  This coherent motion displays
in its own right a number of interesting features, which
we shall describe after having sketched the general formalism.

First we need to define the microscopic model underlying
the density matrix calculation.  The model Hamiltonian is
\begin{eqnarray}\label{TBHam}
H= &&\sum_n\bigl[(\Delta_0^a+neEd)a_n^{\dagger}a_n+
(\Delta_0^b+neEd)b_n^{\dagger}b_n \nonumber\\
&& -{\Delta_1^a\over 4}(a_{n+1}^{\dagger}a_n+a_n^{\dagger}a_{n+1})
+{\Delta_1^b\over 4}(b_{n+1}^{\dagger}b_n+b_n^{\dagger}b_{n+1})
\nonumber\\
&& +eE R( a_n^{\dagger}b_n+b_n^{\dagger}a_n)\bigr ]\;.
\end{eqnarray}
The integers $n$ label the lattice sites and the operators
$a$ and $b$ refer to electrons in the two
minibands; the first two terms give the (field-dependent) site-energies,
the next two describe site-to-site hopping, and the last one
is the term responsible for the interband transfer.  The
overlap matrix-element $R$ is model-dependent and may
depend on time through its momentum argument; we take it as
a constant corresponding to a Kronig-Penney model.
At zero applied field
(\ref{TBHam}) leads to two minibands, $\epsilon^{a,b}(k)
=\Delta_0^{a,b}\mp(\Delta_1^{a,b}/2)\cos(kd)$, while
at finite static fields the spectrum  consists
of two interpenetrating Stark-ladders \cite{Fukuyama}:
$\epsilon^{a,b}_m=\Delta_0^a+eEdr^{a,b}-meEd$.
The numbers $r^{a,b}(E)$ must be determined numerically, and
results of such a calculation are shown in Fig.\ \ref{fig1}.
It is important to notice that for certain field values the two
levels come very close to each other; as we shall show below
this leads to profound
effects on Zener-tunneling.

Let us next consider the equation of motion for the
density matrix for $a$- and $b$-electrons. With accelerated
Bloch states as the basis \cite{Krieger},
the diagonal elements of the density matrix give the
electron density at a $k$-point following the semiclassical trajectory
in reciprocal space,
$k(t)=K-(e/\hbar)\int_0^t E(t')\,dt'$.
We assume equilibrium at $t=0$, when the
density matrix is diagonal. Defining
$\rho_{\pm}(K,t)=\rho^a_K(t)\pm\rho^b_K(t)$, we find
the following equations of motion \cite{long}:
\begin{eqnarray}\label{eqnofmot}
{\dot\rho}_+(K,t)  && = 0\;,\\
{\dot\rho}_-(K,t)  && = -{{Re}}\left\{
h(K,t)\int_0^t dt' h^*(K,t')\rho_-(K,t')\right\}\;,
\end{eqnarray}
where
\begin{equation}\label{defh}
h(K,t)=2{e\over\hbar}E(t)R\exp\left[-\frac{i}{\hbar}\int_0^t
\Delta\epsilon(k(t'))\,dt'\right]\equiv u e^{i\phi}\;.
\end{equation}
In the above equations we have
defined $\Delta\epsilon(k)=\epsilon^b(k)-\epsilon^a(k)$,
and assumed that the intraband couplings are
identical \cite{intraband}.

Eqs.(\ref{eqnofmot}-\ref{defh}) require several comments.
If the interband coupling is turned off, they reduce
to the normal collisionless Boltzmann equation (for two minibands), and
thus contain, as special cases, the following standard
results:
(i) for static field one finds standard Bloch oscillations; and
(ii) a harmonic $E$-field leads to the
analog with Josephson effect of Ref.\cite{Ignatov}, and in
particular, reproduces the band collapse discussed by Holthaus
\cite{Holthaus}.  Note that only $\rho_-$
is affected by interband transitions, while $\rho_+$, which
fixes the particle density, is a constant of motion.

By differentiating the equation of motion for $\rho_-$ with respect
to time, one finds \cite{long}
\begin{equation}\label{rho..}
\ddot{\rho}_-  - {\dot{u}\over u}\dot{\rho}_- +
u^2\rho_-
+u\dot{\phi}\int_0^t dt'{\dot{\phi}\dot{\rho}_-\over u}
=0\;.
\end{equation}

Eq.(\ref{rho..}) forms the basis of our analysis, and the rest of this
paper will describe the numerical results obtained from it
under a number of specific physical conditions.

\paragraph{Steady fields.}  In this case the dc-field is turned on
abruptly at $t=0$. From Eq.(\ref{defh}) $u$ is time-independent, and
Eq.(\ref{rho..}) can be reduced to an ordinary third order
differential equation:
\begin{equation}\label{rho...}
\dot{\phi}\overdots{\rho}_- -\ddot{\phi}\ddot{\rho}_-
+\dot{\phi}(u^2+\dot{\phi}^2)\dot{\rho}_- -u^2\ddot{\phi}\rho_- =0\;.
\end{equation}
The accompanying initial conditions are $\rho_-(0)=
\rho^0_-$, $\dot{\rho}_-(0)=0$, and $\ddot{\rho}_-(0)=-u^2\rho_-(0)$.
Thus, the initial values are determined by the miniband parameters,
the $k$-point in the Brillouin zone
and the temperature (which enters through $\rho_-(0)$).  From
Eq.(\ref{rho...}) it follows that $\rho_-(K,0)$ can be chosen equal to
$1$ without loss of generality.

Fig.\ \ref{fig2} displays a typical time-dependence of $\rho_-$, obtained
from Eq.(\ref{rho...}) by numerical integration.  One observes
a very sensitive behavior with respect to variations of the
applied field: for certain field values an 'inversion' takes
place: $\rho_-$ reaches $-1$, which is the negative of its
initial value, while for other nearby field values $\rho_-$
stays close to its initial value.  This behavior can be understood
by examining Fig.\ \ref{fig1}: the two energy levels are very
close to each other for certain electric field strengths, and
in the corresponding neighborhoods
a strongly enhanced band-to-band transfer takes place.  This
situation is quite different from what one expects from
simple Zener tunneling theory \cite{Zener}: there the tunneling rate is
a monotonic function of the applied field.  The situation is
summarized in Fig.\ \ref{fig3}, where we plot these Zener resonances
as a function of the applied field.  One can assign an index
to the resonances: the resonance at the highest field (which
corresponds approximately to aligning the levels at adjacent quantum wells),
is called the first resonance, the next highest the second
resonance (the case of Fig.\ \ref{fig2}), and so forth.
Adopting this
numbering scheme we observe that in the low field
limit $E^{(n)}\approx\Delta^{ba}/ned$ (here $\Delta^{ba}=\Delta^b_0
-\Delta^a_0$).

Further insight into the physical meaning of the various oscillations
of Fig.\ \ref{fig2} can be obtained by considering the Fourier
transform of $\rho_-$.  Let us first try to establish a qualitative
picture of what to expect.  The initial state of the system is described
by some wave-function, say $|\Psi(0)\rangle$, which is not an eigenstate
after the field has been turned on.  However, it can be expanded in terms of
the eigenstates: $|\Psi(0)\rangle=\sum_n c_n\psi_n^a + d_n \psi_n^b$.
Since $\rho(t)=|\Psi(t)\rangle\langle\Psi(t)|$,
and each eigenstate evolves according to
$\psi_n^{a,b}(t)=\exp(-i\epsilon_n^{a,b}t)\psi_n^{a,b}$,
large
Fourier components in $\rho_-$ are expected to occur at $\hbar\omega=m eEd$
and $\hbar\omega=\pm
eEd(r^a-r^b)+m eEd$\cite{explain}.  This expectation
is fully born out by the numerical evaluation of $\rho_-(\omega)$.
Fig.\ \ref{fig4} shows the results of the two independent calculations:
the continuous lines are obtained based on Fig.\ \ref{fig1}, while
the asterisks come from the Fourier transform of $\rho_-(t)$.  Naturally,
the more laborious calculation based on $\rho_-(t)$ contains also more
information: the magnitudes of the Fourier components
are needed in the evaluation of other
physical quantities, such as the current, which will be addressed
elsewhere \cite{long}.

It is also of interest to examine the effect of varying the superlattice
parameters.  Fig.\ \ref{fig5} shows the time-dependence of
$\rho_-(K=0)$ when the field is tuned to the eighth Zener resonance,
and we have increased the bandwidths and interband coupling.
A distinctive set of stable plateaus
has developed. The transitions
between the plateaus occur at instants $t=\frac{1}{2}T_B, \frac{3}{2}T_B,
\frac{5}{2}T_B...$ after the field was turned on.  Thus, the life-time
of a plateau is (approximately) equal to the Bloch period, and transitions
occur every time the $k$-point reaches the Brillouin zone edge \cite{first}.
This behavior is generic to the {\it low field regime}, $E\leq\Delta^{ba}/ed$,
and we can qualitatively understand features in the time-dependence
of $\rho_-(t)$ by considering the semiclassical motion of a $k$-point
between the extrema of the Brillouin zone; transitions to the other
miniband occur mainly at zone-edges, where the energy separation between
the minibands is at minimum.

We can also understand the {\it number} of oscillations on a
given plateau by examining Eq.(\ref{rho...}) under some simplifying
assumptions. In particular, if we assume that
$\Delta\epsilon$ has a weak time-dependence, it is
easy to solve (\ref{rho...}) analytically. The solution
suggests defining a
'local' time-dependent frequency for a general, but sufficiently slowly
varying
$\Delta\epsilon$ to be
$\omega_l^2(t)=\omega_c^2+\omega_d^2(k(t))$. Here
$\omega_c=2(|R|/d)\omega_B$ and $\hbar\omega_d(k)=\Delta\epsilon(k)$.
Thus, in the low field limit we can identify the number of oscillations
$N_{\mathrm{osc}}\equiv\langle \omega_l(t) \rangle/\omega_B =
\Delta^{ba}/\hbar\omega_B$,
where the time-average was calculated over the Bloch period $T_B$.
Consequently, at the $n$:th Zener resonance, we find
$N_{\mathrm{osc}}=n$ .
In Fig.\ \ref{fig2} one can
distinguish two periods of oscillation
in any of the plateaus (even though the plateaus are not very clearly
resolved for this particular set of parameters), while Fig.\ \ref{fig5}
clearly shows eight periods of oscillation within a plateau.

In the {\it high field regime},
$E>\Delta^{ba}/ed$, the situation differs drastically
from semiclassical expectations: the plateaus
vanish, and we find from Eq.(\ref {rho...}), both numerically
and analytically, that
$\rho_-(t)$ oscillates between -1 and +1 with a single frequency
$\omega_c$.

\paragraph{Alternating fields.}  In the case of a temporally
periodic driving field one can make a close parallel to the treatment
of a spatially periodic potential. Given a Hamiltonian with
$H(t+T)=H(t)$, the starting point is Floquet's theorem applied to the
Schr\"{o}dinger equation. This gives wave-functions of the form
\begin{equation}
\psi_{\epsilon}=\exp(-i\epsilon t)u_{\epsilon}(t)\;,
\end{equation}
where $u_{\epsilon}(t+T)=u_{\epsilon}(t)$. We have introduced the
{\it quasienergy} $\epsilon$, which is defined modulo $\omega_{ac}=2\pi/T$,
leading in a natural way to the definition of a
{\it quasienergy Brillouin zone}. The functions
$\psi_{\epsilon}$ are eigenstates to
$S=H-i\hbar\frac{\partial}{\partial t}$, {\it i.e.} they are
given by the temporal
Bloch theorem applied to $S$.
These considerations have been employed by Holthaus \cite{Holthaus}
in his analysis of the one-miniband case, and we now wish to
extend these concepts to two minibands in the context of
the Hamiltonian (\ref{TBHam}).  The quasienergy spectrum
for an applied field of the form $E(t)=E_1\cos \omega_{ac}t$ is shown
in Fig.\ \ref{fig6} \cite{Niu}.
For vanishing band coupling the band collapses occur
whenever $eE_1d/\hbar\omega_{ac}$ equals a root of the
zeroth Bessel function (the corresponding field strenghts
are indicated by vertical lines in Fig.\ \ref{fig6}),
and we observe that the two-band
model displays strict band collapse at only one of these
roots. However, there is a clear tendency towards
band-width narrowing at the other roots.  Thus, band collapse seems to
be a generic feature, and not a special feature of the previously
studied one-band models.

The next step in the analysis is to solve the equation of
motion for $\rho_-$ for the time-dependent case; one must
now use (\ref{rho..}) as a starting point.
One can perform a completely analogous construction as
is done in Fig.\ \ref{fig4}; now the quasi-energy differences
(modulo $\omega_{ac}$) correspond to the leading Fourier components
of $\rho_-(\omega)$.
There is an important difference
between the static and time-dependent cases \cite{long}:
in the static case the energy spectrum does not depend on
how the field reaches the final value, while in the time-dependent
case the quasienergy bands are translated in reciprocal space by
an amount which is the negative of the semiclassical change in crystal momentum
during the turn-on period. The existence of the corresponding frequencies
in the Fourier spectra of $\rho_-$ can also be directly derived from the
equation of motion for $\rho_-$ \cite{long}.
In the low field limit, $E_1\ll\Delta^{ba}/ed$,
we find that interband tunneling is quenched, and the
time-dependence is described by the semiclassical
equations of motion.

In summary, we have presented a time-dependent formulation
of transport in superlattices.  We have found that the dynamics
of the two-band model can show, in addition to conventional Bloch
oscillations, significant additional structure: Zener resonances,
stable plateaus, and band collapses.

\begin{figure}
\caption{Energy spectrum for a two-band
superlattice as a function of applied dc-field; continuous line
corresponds to Eq.(\protect{\ref{TBHam}}), while asterisks represent a model
with no interband coupling ($R=0$).
The superlattice parameters  are $d=10$nm, $\Delta_1^a=0.8 \times 10^{-2}$eV,
$\Delta_1^b=0.92\times 10^{-2}$eV, $\Delta^{ba}\equiv \Delta_0^b-\Delta_0^a
=2.0 \times 10^{-2}$eV, and $R=-{16\over 9\pi^2}d$.}
\label{fig1}
\end{figure}

\begin{figure}
\caption{Time-dependence of $\rho_-(K=0)$ for three different field values,
top: $eE=0.9\times 10^6$;
middle: $1.1\times 10^6$; and bottom: $1.02\times 10^6$ eV/m, respectively.
For clarity, we have shifted the top and middle curve upwards
by $0.4$ and $0.2$,
respectively.
The superlattice
parameters are as in Fig.\ \protect{\ref{fig1}}, and the unit on the
time axis is $10^3\hbar$/eV$\simeq 4.14$ps.}
\label{fig2}
\end{figure}

\begin{figure}
\caption{The negative of $\rho_-^{\mathrm{min}}$ as a
function of applied field. If
$-\rho_-^{\mathrm{min}}\simeq 1$, a strong band-to-band transfer is taking
place
('Zener resonance').  The dashed lines mark where the Wannier-Stark-
ladder separation has local minima.}
\label{fig3}
\end{figure}

\begin{figure}
\caption{Fourier spectrum of $\rho_-(K=0,t)$.  A dot corresponds to each
significant peak in the Fourier spectrum and the
continuous lines are energy differences
between the interpenetrating Stark ladders.}
\label{fig4}
\end{figure}

\begin{figure}
\caption{Time-dependence of $\rho_-(K=0)$ for superlattice parameters
$\Delta_1^a=\Delta_1^b=1.8\times10^{-2}$eV,
$\Delta^{ba}=2.0\times10^{-2}$eV, and $R=-0.9d$.
Units for the time-axis are as in Fig 2, and the field
is $E=0.232\times10^6$V/m, corresponding to the eighth Zener resonance.}
\label{fig5}
\end{figure}

\begin{figure}
\caption{Quasienergy spectrum for the superlattice of
Fig. 1.
We display the field-dependence of 17 $k$-points, which are evenly
distributed in the positive half of the Brillouin zone, $k\in [0,\pi/d]$
(due to evenness of the quasienergy spectrum negative $k$'s would
be redundant).
The vertical lines indicate where
the band collapse would occur for non-interacting minibands.
The modulation frequency is
$\omega_{\mathrm{ac}}=1.5\times10^{-2}$eV/$\hbar$, and the
quasienergies are given in units of $\omega_{\mathrm{ac}}$.}
\label{fig6}
\end{figure}


\begin{references}

\bibitem{Bloch}
F. Bloch, Z. Physik {\bf 52}, 555 (1928).

\bibitem{Zener}
C. Zener, Proc. Roy. Soc. {\bf A145}, 523 (1934).

\bibitem{EsakiTsu}
L. Esaki and R. Tsu, IBM Journal of Res. Develop. {\bf 14}, 61 (1970).

\bibitem{Waschke}
C. Waschke, H. G. Roskos, R. Schwedler, K. Leo, H. Kurz, and K. K{\"o}hler,
Phys. Rev. Lett. {\bf 70}, 3318 (1993).  This paper also contains
references to other relevant experimental work.

\bibitem{Holthaus}
M. Holthaus, Phys. Rev. Lett. {\bf 69}, 351 (1992); Z.Phys. B - Cond. Matter
{\bf 89}, 251 (1992).

\bibitem{Dunlap}
A. A. Ignatov and Y. A. Romanov, phys. stat. sol. (b) {\bf 73}, 327 (1976);
D. H. Dunlap, and V. M. Kenkre, Phys. Rev. B {\bf 34}, 3625 (1986).

\bibitem{Ignatov}
A. A. Ignatov, K. F. Renk, and E. P. Dodin, Phys. Rev. Lett. {\bf 70},
1996 (1993).

\bibitem{Meier}
T. Meier, G. von Plessen, P. Thomas, and S. W. Koch, Phys. Rev. Lett.
{\bf 73}, 902 (1994).

\bibitem{Holt2B} See, however, the very recent paper by M. Holthaus and
D. W. Hone, Phys. Rev. B {\bf 49}, 16605 (1994), which analyzes a finite
multiband superlattice in an ac field.

\bibitem{Fukuyama}
H. Fukuyama, R. A. Bari, and H. C. Fogedby, Phys. Rev. B {\bf 8},
5579 (1973).

\bibitem{Krieger}
J. B. Krieger and G. J. Iafrate, Phys. Rev. B {\bf 33}, 5394 (1986);
{\it ibid.} {\bf 35}, 9644 (1987); {\it ibid.} {\bf 38}, 6324 (1988).

\bibitem{long}
J. Rotvig, A. P. Jauho, and H. Smith (unpublished).

\bibitem{intraband} Explicitly, we assume that $R_{11}=R_{22}$,
where $R_{nn}(k)={i\over d}\int_{-d/2}^{d/2}dx u_{n,k}^*(x)\nabla_k
u_{n,k}(x)$.  This assumption is valid e.g. for the Kronig-Penney
model.

\bibitem{explain} Here we used the known energy spectrum of
the biased superlattice \cite{Fukuyama}.

\bibitem{first} Since $K=0$ in Fig.\ \ref{fig5}, which corresponds
to the center of the Brillouin zone, the first transition occurs
at $t=T_B/2$.

\bibitem{Niu} In our numerical calculation we used the method
suggested by X. - G. Zhao, R. Jahnke, and Q. Niu (unpublished).

\end{references}
\end{document}